\documentclass[%
 aip,
 amsmath,amssymb,
 reprint,%
]{revtex4-1}

\usepackage{graphicx}
\usepackage{dcolumn}
\usepackage{bm}

\usepackage[utf8]{inputenc}
\usepackage[T1]{fontenc}
\usepackage{mathptmx}
\usepackage{etoolbox}

\usepackage{siunitx}
\usepackage{gensymb}

\makeatletter
\def\@email#1#2{%
 \endgroup
 \patchcmd{\titleblock@produce}
  {\frontmatter@RRAPformat}
  {\frontmatter@RRAPformat{\produce@RRAP{*#1\href{mailto:#2}{#2}}}\frontmatter@RRAPformat}
  {}{}
}%
\makeatother
\begin{document}

\preprint{AIP/123-QED}

\title[]{Impact of magnetic anisotropy on the magnon Hanle effect in $\alpha$-Fe$_2$O$_3$}

\author{M. Scheufele}
\altaffiliation[]{Author to whom correspondence should be addressed: monika.scheufele@wmi.badw.de}
 \affiliation{Walther-Mei{\ss}ner-Institut, Bayerische Akademie der Wissenschaften, 85748 Garching, Germany}
 \affiliation{Technische Universit\"{a}t M\"{u}nchen, TUM School of Natural Sciences, Physics Department, 85748 Garching, Germany}
 
\author{J. G{\"u}ckelhorn}%
\affiliation{Walther-Mei{\ss}ner-Institut, Bayerische Akademie der Wissenschaften, 85748 Garching, Germany}
\affiliation{Technische Universit\"{a}t M\"{u}nchen, TUM School of Natural Sciences, Physics Department, 85748 Garching, Germany}

\author{M. Opel}%
\affiliation{Walther-Mei{\ss}ner-Institut, Bayerische Akademie der Wissenschaften, 85748 Garching, Germany}

\author{A. Kamra}
\affiliation{Condensed Matter Physics Center (IFIMAC) and Departamento de F\'{i}sica Te\'{o}rica de la Materia Condensada, Universidad Aut\'{o}noma de Madrid, 28049 Madrid, Spain}

\author{H. Huebl}%
\affiliation{Walther-Mei{\ss}ner-Institut, Bayerische Akademie der Wissenschaften, 85748 Garching, Germany}
\affiliation{Technische Universit\"{a}t M\"{u}nchen, TUM School of Natural Sciences, Physics Department, 85748 Garching, Germany}
\affiliation{Munich Center for Quantum Science and Technology (MCQST), 80799 Munich, Germany}

\author{R. Gross}%
\affiliation{Walther-Mei{\ss}ner-Institut, Bayerische Akademie der Wissenschaften, 85748 Garching, Germany}
\affiliation{Technische Universit\"{a}t M\"{u}nchen, TUM School of Natural Sciences, Physics Department, 85748 Garching, Germany}
\affiliation{Munich Center for Quantum Science and Technology (MCQST), 80799 Munich, Germany}

\author{S. Gepr{\"a}gs}
\affiliation{Walther-Mei{\ss}ner-Institut, Bayerische Akademie der Wissenschaften, 85748 Garching, Germany}

\author{M. Althammer}
\altaffiliation[]{Electronic mail: matthias.althammer@wmi.badw.de}
\affiliation{Walther-Mei{\ss}ner-Institut, Bayerische Akademie der Wissenschaften, 85748 Garching, Germany}
\affiliation{Technische Universit\"{a}t M\"{u}nchen, TUM School of Natural Sciences, Physics Department, 85748 Garching, Germany}

\date{\today}

\begin{abstract}
In easy-plane antiferromagnets, the nature of the elementary excitations of the spin system is captured by the precession of the magnon pseudospin around its equilibrium pseudofield, manifesting itself in the magnon Hanle effect. Here, we investigate the impact of growth-induced changes in the magnetic anisotropy on this effect in the antiferromagnetic insulator \mbox{$\alpha$-Fe$_2$O$_3$} (hematite). To this end, we compare the structural, magnetic, and magnon-based spin transport properties of \mbox{$\alpha$-Fe$_2$O$_3$} films with different thicknesses grown by pulsed laser deposition in molecular and atomic oxygen atmospheres. While in films grown with molecular oxygen a spin-reorientation transition (Morin transition) is absent down to $\SI{10}{\kelvin}$, we observe a Morin transition for those grown by atomic-oxygen-assisted deposition, indicating a change in magnetic anisotropy. Interestingly, even for a $\SI{19}{\nano \meter}$ thin \mbox{$\alpha$-Fe$_2$O$_3$} film grown with atomic oxygen we still detect a Morin transition at $\SI{125}{K}$. We characterize the magnon Hanle effect in these \mbox{$\alpha$-Fe$_2$O$_3$} films via all-electrical magnon transport measurements. The films grown with atomic oxygen show a markedly different magnon spin signal from those grown in molecular oxygen atmospheres. Most importantly, the maximum magnon Hanle signal is significantly enhanced and the Hanle peak is shifted to lower magnetic field values for films grown with atomic oxygen. These observations suggest a change of magnetic anisotropy for \mbox{$\alpha$-Fe$_2$O$_3$} films fabricated by atomic-oxygen-assisted deposition resulting in an increased oxygen content in these films. Our findings provide new insights into the possibility to fine-tune the magnetic anisotropy in \mbox{$\alpha$-Fe$_2$O$_3$} and thereby to engineer the magnon Hanle effect.
\end{abstract}


\maketitle

\section{\label{sec:intro}Introduction}
The field of magnonics, utilizing the quantized spin excitations of magnetically ordered systems, i.e.\,magnons, offers a variety of interesting opportunities such as information processing with bosons and magnon-based computing approaches \cite{Chumak2015,Nakata2017,Althammer2021}. Antiferromagnets are particularly appealing for magnonic logic devices due to their very fast magnetization dynamics in the THz regime and their robustness against external magnetic fields \cite{Olejník2018,Vaidya2020,Li2020,Jungwirth2016,Baltz2018}. Antiferromagnetic magnons come in pairs with opposite precession chiralities of the N\'{e}el order vector and thus opposite pseudospin\cite{Kamra2020,Wimmer2020}. In the case of easy-plane antiferromagnets, linearly polarized spin waves with zero effective spin, which can be viewed as an equal superposition of basis states with opposite chirality, form the appropriate eigenmodes of the spin system \cite{Kawano2019,Daniels2018,Liensberger2019,Cheng2016,Shen2020}. 
The manipulation and read-out of information encoded into antiferromagnetic magnons is demanding due to the vanishing stray fields of antiferromagnets. However, it has been demonstrated that this can be achieved by all-electrical magnon transport \cite{Cornelissen2015,Gonnenwein2015,Cornelissen2016,Velez2016,Zhang2012,Li2016}. There, the electrical injection and detection of magnonic spin transport is realized via two heavy-metal electrodes adjacent to the antiferromagnetic insulator by making use of the spin Hall and inverse spin Hall effect \cite{Hirsch1999,Sinova2015,Lebrun2018,Valenzuela2006,Saitoh2006}. In such experiments the magnon Hanle effect, the analog of the electron Hanle effect, has been demonstrated in the easy-plane antiferromagnetic insulator \mbox{$\alpha$-Fe$_2$O$_3$}. This has been achieved by a coherent control of the magnon pseudospin and long-distance magnon-based spin propagation in \mbox{$\alpha$-Fe$_2$O$_3$} \cite{Lebrun2018}.

Bulk hematite (\mbox{$\alpha$-Fe$_2$O$_3$}) is an antiferromagnetic insulator (AFI) below the N\'{e}el temperature $T_\mathrm{N}=\SI{953}{\kelvin}$ and exhibits a spin-reorientation transition at the Morin transition temperature of $T_\mathrm{M}=\SI{263}{\kelvin}$ \cite{Morin1950}. In the temperature range $T_\mathrm{M} < T < T_\mathrm{N}$, the magnetic moments are parallel to the magnetically easy \mbox{$\alpha$-Fe$_2$O$_3$} ($0001$)-plane as described by a uniaxial magnetic anisotropy along the [$0001$]-trigonal axis. Additionally, a finite spin canting is induced by the Dzyaloshinskii–Moriya interaction (DMI) resulting in a net magnetization of $\SI{2.5}{\kilo\ampere\per\meter}$ perpendicular to the [0001] hard axis at room temperature \cite{Morin1950,Morrish1995}. The easy-plane anisotropy 
gives rise to the magnon Hanle effect in \mbox{$\alpha$-Fe$_2$O$_3$} described by the precessional motion of antiferromagnetic magnon pseudospin around its equilibrium pseudofield oriented along the $x$-axis (see coordinate system in Fig.\,\ref{fig:hanle} (a)) \cite{Wimmer2020,Kamra2020}. For $T<T_\mathrm{M}$, the uniaxial magnetic anisotropy reverses sign, leading to a reorientation of the magnetic Fe$^{3+}$-moments along the [$0001$]-axis representing now the magnetically easy axis \cite{Morin1950,Besser1967}. In this easy-axis antiferromagnetic configuration, the net magnetization vanishes and one expects the magnon Hanle effect to disappear. This conjecture, however, is difficult to prove since films with small thickness are desirable in all-electrical magnon transport experiments to discern the magnon Hanle signature from the finite spin signal stemming from low energy magnons \cite{Guckelhorn2022}. However, in thin \mbox{$\alpha$-Fe$_2$O$_3$} films with a thickness below $\SI{100}{\nano \meter}$ the Morin transition is usually suppressed as size-related effects change the delicate balance between magnetic-dipolar and uniaxial anisotropy contributions, which determines the Morin transition temperature \cite{Shimomura2015,Artman}. Therefore, the identification of growth conditions for \mbox{$\alpha$-Fe$_2$O$_3$} thin films, which show a Morin transition, are key for confirming the conjecture that an easy plane anisotropy is required for the presence of the magnon Hanle effect. 


In this article, we investigate the crystallographic, magnetic and magnon transport properties of epitaxial \mbox{$\alpha$-Fe$_2$O$_3$} films with thickness values ranging from $\SI{15}{\nano\meter}$ to $\SI{124}{\nano\meter}$. These films have been grown by pulsed laser deposition using two different growth atmospheres: (i) pure molecular oxygen and (ii) molecular oxygen with additional atomic oxygen. We observe a finite Morin transition temperature $T_\mathrm{M}$ in \mbox{$\alpha$-Fe$_2$O$_3$} films fabricated by atomic-oxygen-assisted deposition, while a Morin transition is absent in \mbox{$\alpha$-Fe$_2$O$_3$} films grown in molecular oxygen atmosphere. This suggests a modification of the magnetic anisotropy in \mbox{$\alpha$-Fe$_2$O$_3$} by changing the oxygen content.
In all-electrical magnon transport measurements, the atomic-oxygen assisted growth of \mbox{$\alpha$-Fe$_2$O$_3$} yields a modification of the antiferromagnetic magnon pseudospin dynamics. Within the scope of the magnon Hanle effect, this manifests itself in an enhanced magnon spin signal amplitude and a shift of the Hanle peak towards smaller magnetic field magnitudes. 

\section{Sample fabrication and characterization}
\label{sec:fabrication}
Epitaxial \mbox{$\alpha$-Fe$_2$O$_3$} films are grown via pulsed laser deposition on ($0001$)-oriented, single crystalline sapphire (Al$_2$O$_3$) substrates. The growth process is carried out in an atmosphere of molecular oxygen with a partial pressure of $\SI{25}{\micro \bar}$, while the substrate temperature is kept at $\SI{320}{\celsius}$. The laser fluence at the polycrystalline \mbox{$\alpha$-Fe$_2$O$_3$} target is $\SI{2.5}{\joule\per\centi\meter\squared}$ and the pulse repetition rate is set to $\SI{2}{\hertz}$ \cite{Fischer2020}. To reduce the formation of oxygen vacancies, we fabricated a second set of \mbox{$\alpha$-Fe$_2$O$_3$} films utilizing an atomic oxygen RF source (AOS) at a power of $\SI{400}{\watt}$, where we add atomic oxygen to the deposition atmosphere, while all other deposition parameters are kept constant. In the following, we differentiate between \mbox{NAOS-Fe$_2$O$_3$} films (no AOS used during deposition) and \mbox{AOS-Fe$_2$O$_3$} films (AOS used during deposition). The thicknesses of the \mbox{$\alpha$-Fe$_2$O$_3$} films discussed in this article range from $\SI{15}{\nano \meter}$ to $\SI{124}{\nano \meter}$ to investigate the differences between NAOS- and \mbox{AOS-Fe$_2$O$_3$} for both thin and thick films. Hereby, we consider \mbox{$\alpha$-Fe$_2$O$_3$} films as thin (thick), if the film thickness $t_\mathrm{m}$ is comparable to (much larger than) the thermal magnon wavelength $l_\mathrm{th}$, which is typically much smaller than the magnon spin decay length. The thin and thick film limits and their influence on the magnon Hanle signal have been studied in detail in our previous work \cite{Guckelhorn2022}. The thickness values of the discussed \mbox{$\alpha$-Fe$_2$O$_3$} films are summarized in Table \ref{tab:table}.
\begin{table}
\caption{\label{tab:table} Sample overview. The thicknesses are specified for the investigated \mbox{$\alpha$-Fe$_2$O$_3$} films analyzed by different measurement methods (see footnotes).}
\begin{ruledtabular}
\begin{tabular}{cll}
Atomic oxygen source used&thin $\alpha$-Fe$_2$O$_3$&thick $\alpha$-Fe$_2$O$_3$\\
\hline
No & $\SI{15}{\nano \meter}$\footnote{2$\theta$-$\omega$ X-ray scan.}$^\text{,d}$ & $\SI{103}{\nano \meter}^\text{a,c,d}$\\
No & $\SI{25}{\nano \meter}$\footnote{2$\theta$-$\omega$ X-ray scan and reciprocal space map (see supplementary material).}$^\text{,}$\footnote{SQUID magnetometry.} & $\SI{124}{\nano \meter}^\text{b}$\\
Yes & $\SI{19}{\nano \meter}^\text{a,b,c}$\footnote{All-electrical diffusive magnon transport.} & $\SI{89}{\nano \meter}^\text{a,b,c,d}$\\
\end{tabular}
\end{ruledtabular}
\end{table}

\begin{figure}
\includegraphics{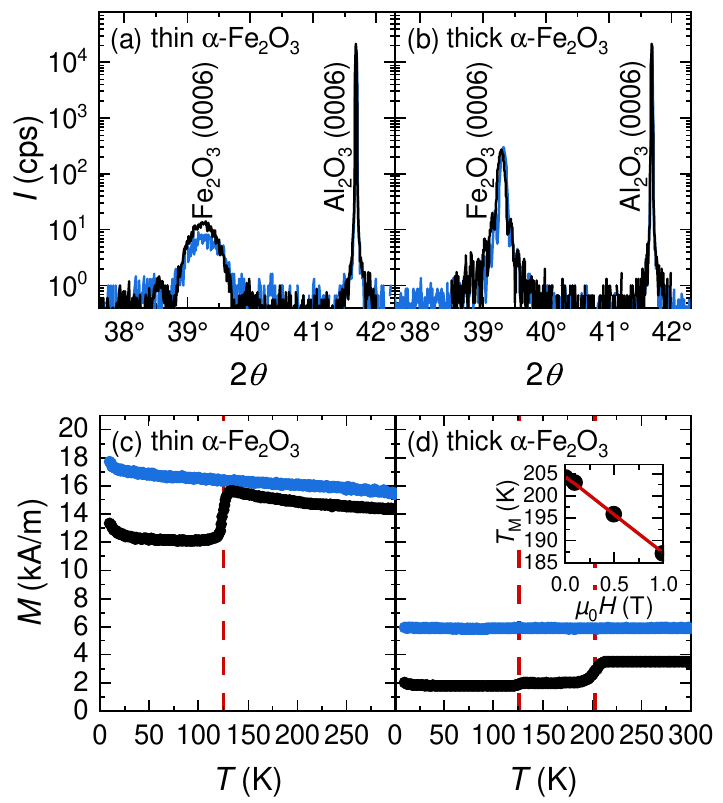}
\caption{\label{fig:xrd} Structural and magnetic properties of (a), (c) thin and (b), (d) thick \mbox{$\alpha$-Fe$_2$O$_3$} films measured by high-resolution X-ray diffraction (upper panels) and SQUID magnetometry (lower panels). Blue lines indicate \mbox{$\alpha$-Fe$_2$O$_3$} films deposited in molecular oxygen atmosphere (\mbox{NAOS-Fe$_2$O$_3$} films) and black lines \mbox{$\alpha$-Fe$_2$O$_3$} films fabricated by atomic-oxygen-assisted deposition (\mbox{AOS-Fe$_2$O$_3$} films). The $2\theta$-$\omega$ scans are performed around the \mbox{$\alpha$-Fe$_2$O$_3$} ($0006$) film and the Al$_2$O$_3$ ($0006$) substrate reflections. The \mbox{$\alpha$-Fe$_2$O$_3$} films have a thickness of (a) $\SI{15}{\nano \meter}$ (blue) and $\SI{19}{\nano \meter}$ (black) as well as (b) $\SI{103}{\nano \meter}$ (blue) and $\SI{89}{\nano \meter}$ (black). For the temperature-dependent magnetization measurements in (c) and (d), the same \mbox{$\alpha$-Fe$_2$O$_3$} films were investigated except for the thin \mbox{NAOS-Fe$_2$O$_3$} film. Here, a $\SI{25}{\nano \meter}$ thin \mbox{$\alpha$-Fe$_2$O$_3$} film is utilized. The red, dashed vertical lines indicate the Morin transitions visible for the \mbox{AOS-Fe$_2$O$_3$} films. All SQUID magnetometry measurements are conducted while heating the samples in an in-plane magnetic field of $\mu_0H=\SI{100}{\milli \tesla}$ after cooling down in zero field. The inset in (d) displays the extracted Morin transition temperature $T_\text{M}$ as a function of the measurement field $\mu_0H$ together with a linear fit in red.}
\end{figure}

We analyze the structural properties of our \mbox{$\alpha$-Fe$_2$O$_3$} films by high-resolution X-ray diffraction (HR-XRD) measurements. The corresponding $2\theta$-$\omega$ scans around the \mbox{$\alpha$-Fe$_2$O$_3$} ($0006$) film and Al$_2$O$_3$ ($0006$) substrate reflections are depicted in Fig.\,\ref{fig:xrd} (a), (b) for thin and thick \mbox{$\alpha$-Fe$_2$O$_3$} films, respectively. Thereby, we observe no secondary crystalline phases. Finite thickness fringes around the \mbox{$\alpha$-Fe$_2$O$_3$} ($0006$) reflections indicate a coherent growth and thus a good crystalline quality of the \mbox{$\alpha$-Fe$_2$O$_3$} films\cite{Geprags2020}. Moreover, the ($0006$) reflections of the \mbox{AOS-Fe$_2$O$_3$} films (black lines) are slightly shifted towards smaller $2\theta$-values with respect to the ones of the \mbox{NAOS-Fe$_2$O$_3$} films (blue lines). This indicates an increase in the out-of-plane lattice constant $c$, when adding atomic oxygen to the fabrication process and thus reducing the oxygen deficiency in \mbox{$\alpha$-Fe$_2$O$_3$}. Reciprocal space maps (RSMs) around the asymmetric \mbox{$\alpha$-Fe$_2$O$_3$} ($10\bar{1}\,10$) and Al$_2$O$_3$ ($10\bar{1}\,10$) reflections also confirm a slight reduction in $c$ for AOS-Fe$_2$O$_3$ films, while the in-plane lattice constant $a$ remains nearly constant. It should be mentioned that the changes in $c$ are small and within the range of the measurement uncertainty. The extracted lattice parameters indicate a nearly relaxed growth of \mbox{$\alpha$-Fe$_2$O$_3$} on the Al$_2$O$_3$ substrates for all samples investigated. We obtain comparable values of the small, but finite epitaxial in-plane strain $\epsilon_\mathrm{xx}$ for NAOS- and \mbox{AOS-Fe$_2$O$_3$} films suggesting that the value of $\epsilon_\mathrm{xx}$ is independent of the growth method. Furthermore, we do not observe any indication of an additional, strained \mbox{$\alpha$-Fe$_2$O$_3$} layer near the film-substrate interface due to clamping effects as stated in a recent publication \cite{Wittmann2022}. A detailed discussion on the RSMs and the extracted lattice constants is given in the supplementary material (SM). 

To identify possible magnetic phase transitions in our \mbox{$\alpha$-Fe$_2$O$_3$} films, we performed superconducting quantum interference device (SQUID) magnetometry. After cooling the samples down to $\SI{10}{\kelvin}$ in zero field, we measured the magnetization $M$ as a function of temperature $T$ at a fixed magnetic field of $\mu_0H = \SI{100}{\milli \tesla}$ applied in the film plane. The corresponding results are presented in Fig.\,\ref{fig:xrd} (c) and (d), where a temperature-independent background signal was subtracted beforehand. This background signal is a linear function of $\mu_0H$, which mainly stems from the diamagnetic Al$_2$O$_3$ substrate, but also from an increased sublattice canting in \mbox{$\alpha$-Fe$_2$O$_3$} with increasing $\mu_0H$\cite{Geprags2020}. The \mbox{NAOS-Fe$_2$O$_3$} films exhibit no phase transitions over the whole temperature range from $\SI{300}{\kelvin}$ down to $\SI{10}{\kelvin}$ and thus remain in the magnetically ($0001$)-easy plane phase with a finite net magnetization induced by DMI.
Generally, the shift of the Morin transition towards lower temperatures or its complete absence in \mbox{$\alpha$-Fe$_2$O$_3$} films can be attributed to strain-induced changes of the magnetic anisotropy \cite{Han2020,Mibu2017,Park2013}. However, as discussed above, all investigated \mbox{$\alpha$-Fe$_2$O$_3$} films are nearly relaxed and therefore exhibit only small in-plane strain values (see SM). At low temperatures, some of the \mbox{$\alpha$-Fe$_2$O$_3$} films show a small increase in $M$ (cf. Fig.\,\ref{fig:xrd} (c)), which is most probably caused by paramagnetic moments within the Al$_2$O$_3$ substrate.

The \mbox{AOS-Fe$_2$O$_3$} films, instead, reveal Morin transition temperatures of $T_\mathrm{M}\approx\SI{205}{\kelvin}$ and $T_\mathrm{M}\approx\SI{125}{\kelvin}$ for thick and thin films, respectively, which are found to decrease with increasing applied magnetic field (see inset of Fig.\,\ref{fig:xrd} (d)). As the small in-plane strain is comparable for NAOS- and \mbox{AOS-Fe$_2$O$_3$} films, the absence (appearance) of the Morin transition in NAOS \mbox{(AOS)-Fe$_2$O$_3$} can not be explained by strain-induced effects on the magnetic anisotropy. The atomic-oxygen-assisted deposition could rather lead to a reduction of oxygen vacancies and therefore to a change of the magnetic anisotropy in \mbox{AOS-Fe$_2$O$_3$} films compared to the possibly more oxygen deficient \mbox{NAOS-Fe$_2$O$_3$} films. We note that an oxygen deficiency in the \mbox{$\alpha$-Fe$_2$O$_3$} films results in a partial reduction of Fe$^{3+}$ to Fe$^{2+}$ ions and thus is expected to change the magnetic anisotropy and $T_\mathrm{M}$ \cite{Jani2021}. However, the Morin transition temperature of the \mbox{AOS-Fe$_2$O$_3$} films is still smaller than in bulk crystals, which has also been recently reported for \mbox{$\alpha$-Fe$_2$O$_3$} thin films \cite{Shimomura2015,Mibu2017}. 
Additionally, a second Morin transition at $T_\mathrm{M}^*\approx\SI{125}{\kelvin}$ indicated by the small kink in the $M$($T$) curve can be observed for the thick \mbox{AOS-Fe$_2$O$_3$} film (see  Fig.\,\ref{fig:xrd} (d)). 
This low temperature Morin transition coincides with the single Morin transition in the thin \mbox{AOS-Fe$_2$O$_3$} film and suggests that it originates from a region near the film-substrate interface with a different magnetic anisotropy. 
The detailed origin of the different magnetic anisotropy remains unknown. However, we can speculate that it is related to an increased defect density (e.g.\,oxygen vacancies) at the film-substrate interface, which affects the magnetic properties of our \mbox{$\alpha$-Fe$_2$O$_3$} films\cite{Warschkow2002}.


Although we expect a drop of $M$ to zero for $T<T_\mathrm{M}$ in the easy-axis phase, we observe a surprisingly large magnetization also for $T<\SI{125}{\kelvin}$ for the \mbox{AOS-Fe$_2$O$_3$} films. Since we do not find any indications from HR-XRD for the presence of other iron oxide phases such as ferrimagnetic $\gamma$-Fe$_2$O$_3$ (maghemite) or Fe$_3$O$_4$ (magnetite) that could contribute to the finite magnetization of the samples, we suggest that the spin-reorientation at $T_\mathrm{M}$ is incomplete and leads to co-existing easy-plane and easy-axis phases in our \mbox{AOS-Fe$_2$O$_3$} films below the respective Morin transitions. Notably, the net magnetization is larger in thin than in thick films, which can be induced by changes in the DMI, the exchange coupling or the magnetic anisotropy in \mbox{$\alpha$-Fe$_2$O$_3$}. This observation is interesting and demonstrates that the impact of film thickness on the magnetic properties of the \mbox{$\alpha$-Fe$_2$O$_3$} films has to be taken into account and needs detailed consideration. In total, our SQUID magnetometry measurements suggest a more complex spin structure of our \mbox{$\alpha$-Fe$_2$O$_3$} films proving that an extensive study of the Morin transition can be a powerful tool to investigate the magnetic anisotropy in hematite films.

\section{All-electrical magnon transport}
\label{sec:transport}
\begin{figure}
\includegraphics{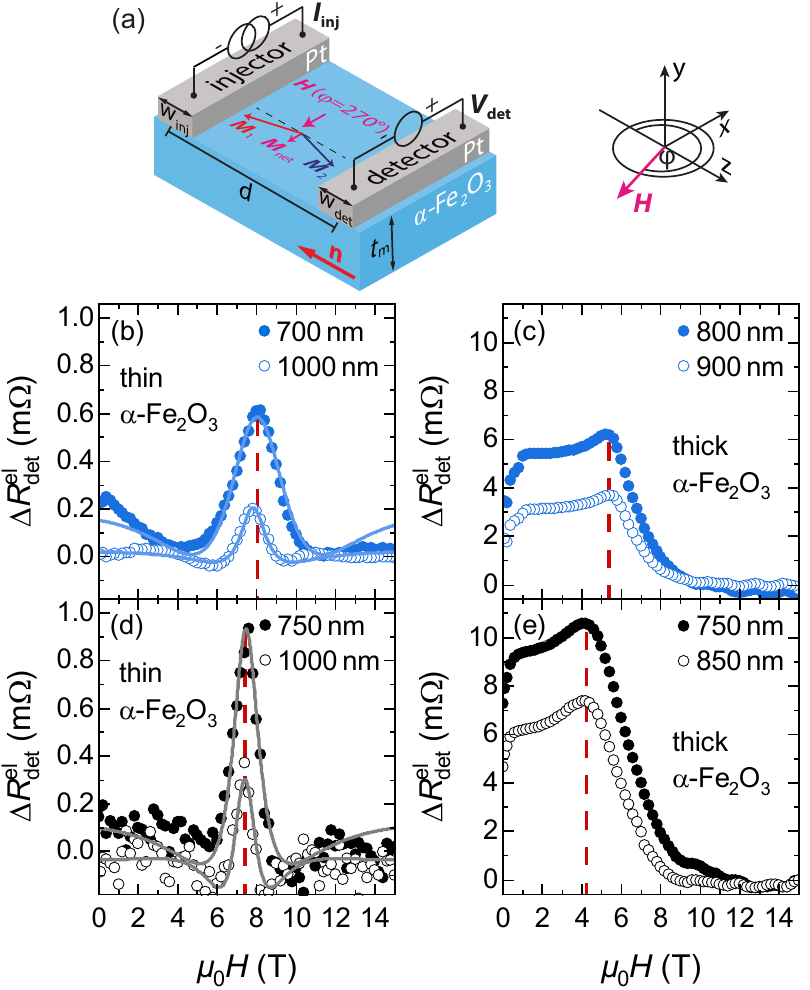}
\caption{\label{fig:hanle} (a) Scheme of the sample configuration consisting of two Pt strips on top of a \mbox{$\alpha$-Fe$_2$O$_3$} film in the magnetic ($0001$) easy-plane phase. The sublattice magnetizations $\pmb{M}_1$ and $\pmb{M}_2$ are slightly canted resulting in a net magnetization $\pmb{M}_\mathrm{net}$. The N\'{e}el order parameter $\pmb{n}\perp\pmb{M}_\mathrm{net}$ is controlled by applying a magnetic field $\pmb{H}$ that is rotated in the film plane by the angle $\varphi$. A charge current is applied at the left Pt-electrode and the magnon spin signal is detected as a voltage at the right Pt-electrode. (b)-(e) Amplitude of the electrically induced magnon spin signal $\Delta R_\text{det}^\text{el}$ as a function of $\mu_0H$ for thin (left panels) as well as for thick \mbox{$\alpha$-Fe$_2$O$_3$} films (right panels). The measurements are conducted at $\SI{200}{\kelvin}$ utilizing different injector-detector separations $d$ (full and open circles). The $\SI{19}{\nano \meter}$ and $\SI{89}{\nano \meter}$ thick \mbox{AOS-Fe$_2$O$_3$} films in (d) and (e) exhibit larger magnon spin signals than the $\SI{15}{\nano \meter}$ and $\SI{103}{\nano \meter}$ thick \mbox{NAOS-Fe$_2$O$_3$} films in (b) and (c). The solid lines in (b) and (d) are fits to Eq.\,\eqref{eq: hanle}. The red, dashed lines indicate the compensation field $\mu_0H_\text{c}$, where $\Delta R_\text{det}^\text{el}$ is maximum.}
\end{figure}

In the following, we study the impact of different magnetic anisotropies induced by different oxygen atmospheres during the deposition of \mbox{$\alpha$-Fe$_2$O$_3$} films on their magnon transport properties. For our experiments, two $\SI{500}{\nano\meter}$ wide and $\SI{5}{\nano\meter}$ thick Pt strip electrodes with different center-to-center distances $d$ (see Fig.\,\ref{fig:hanle} (a)) are patterned on top of the \mbox{$\alpha$-Fe$_2$O$_3$} films via lift-off process using electron-beam lithography and sputter deposition. We apply a DC charge current $I_\mathrm{inj}=\SI{500}{\micro\ampere}$ to one Pt electrode injecting a spin current into the \mbox{$\alpha$-Fe$_2$O$_3$} film via the spin Hall effect (SHE) \cite{Hirsch1999,Sinova2015,Lebrun2018,Valenzuela2006}. The hereby excited diffusive pseudospin magnon current is electrically detected in the second electrode as a voltage signal $V_\mathrm{det}$ via the inverse SHE (iSHE) \cite{Saitoh2006,Sinova2015}. We use the current reversal technique to extract the voltage signal $V_\mathrm{det}^\mathrm{el}$ originating from the SHE excited magnons \cite{Gonnenwein2015,Ganzhorn2016}. For fabrication and measurement details we refer to the SM. The electrically induced magnon spin signal $R_\mathrm{det}^\mathrm{el}=V_\mathrm{det}^\mathrm{el}/I_\mathrm{inj}$ depends on the orientation of the N\'{e}el vector $\pmb{n}$ with respect to the spin polarization $\pmb{s}$ of the injected spin current. The N\'{e}el vector is defined by the sublattice magnetizations $\pmb{M}_1$ and $\pmb{M}_2$ with corresponding saturation magnetizations $M_1$ and $M_2$ resulting in $\pmb{n}=(\pmb{M}_1/M_1-\pmb{M}_2/M_2)/2$. To orient the net magnetization $\pmb{M}_\mathrm{net}=\pmb{M}_1+\pmb{M}_2$ and thus $\pmb{n}\perp\pmb{M}_\mathrm{net}$, we apply a magnetic field $\pmb{H}$ in the film plane. We find $R_\mathrm{det}^\mathrm{el}$ to be maximum (zero) for $\pmb{n}\parallel\pmb{s}$ ($\pmb{n}\perp\pmb{s}$), which is the case for $\pmb{H}$ at $\varphi=\SI{270}{\degree}$ ($\SI{180}{\degree}$) (see coordinate system in Fig.\,\ref{fig:hanle} (a)). 

This allows us to extract the electrically induced magnon spin signal amplitude $\Delta R_\text{det}^\text{el}=R_\text{det}^\text{el}(\varphi=\SI{270}{\degree})-R_\text{det}^\text{el}(\varphi=\SI{180}{\degree})$ that is shown in Fig.\,\ref{fig:hanle} (b)-(e) as a function of $\mu_0H$ for different $d$ (open and full circles). The measurements are conducted at $\SI{200}{\kelvin}$, i.e.\,in the magnetic easy-plane phase of the \mbox{NAOS-Fe$_2$O$_3$} films and of the thin \mbox{AOS-Fe$_2$O$_3$} film ($T_\mathrm{M}\le\SI{125}{\kelvin}$). In case of the thick \mbox{AOS-Fe$_2$O$_3$} film, our SQUID magnetometry results (inset in Fig.~\ref{fig:xrd}(d)) indicate that a magnetic field of $\SI{250}{\milli\tesla}$ is large enough to shift $T_\mathrm{M}$ below $\SI{200}{\kelvin}$ and we maintain the easy-plane phase in the magnon transport experiments. In (b), the thin \mbox{NAOS-Fe$_2$O$_3$} film exhibits the characteristic magnon Hanle curve, which is discussed in detail in our previous works \cite{Wimmer2020,Guckelhorn2022,Guckelhorn2022b}. Note that $\Delta R_\text{det}^\text{el}$ is maximum at the so-called compensation field $\mu_0H_\text{c}$ of about $\SI{8}{\tesla}$ for both values of $d$. At $\mu_0H_\text{c}$, the pseudofield $\omega$ is zero and hence the injected magnons propagate without any pseudospin precession \cite{Wimmer2020}. The compensation field is therefore independent of the injector-detector distance and only depends on the material parameters of the hematite film. Furthermore, $\Delta R_\text{det}^\text{el}$ strongly decreases with increasing $d$, since the number of magnons carrying the spin current is not a conserved quantity, but decays within the magnon spin relaxation time $\tau_\mathrm{m}$. To quantify $\tau_\mathrm{m}$, we fit $\Delta R_\text{det}^\text{el}$ to the detectable $z$-component of the pseudospin chemical potential \cite{Kamra2020}
\begin{equation}
    \label{eq: hanle}
    \mu_\mathrm{sz} = \frac{l_\mathrm{m}j_\mathrm{s0}e^{-\frac{ad}{l_\mathrm{m}}}}{D_\mathrm{m}\chi(a^2+b^2)}\bigg(a\cos{\bigg(\frac{bd}{l_\mathrm{m}}\bigg)}-b\sin{\bigg(\frac{bd}{l_\mathrm{m}}\bigg)}\bigg)
\end{equation}
with $a, b = \sqrt{(\sqrt{1+\omega^2\tau_\mathrm{m}^2}\pm 1)/2}$ and the magnon spin decay length $l_\mathrm{m}=\sqrt{D_\mathrm{m}\tau_\mathrm{m}}$ (see SM for the extracted fit parameters). Here, $j_\mathrm{s0}$ is the magnon spin current density driven by the injector, $D_\mathrm{m}$ is the magnon diffusion constant and $\chi$ is the susceptibility relating the pseudospin density to the pseudospin chemical potential \cite{Kamra2020}. This fitting approach is only applicable for thin hematite films. In (c), the thick \mbox{NAOS-Fe$_2$O$_3$} film reveals an offset in $\Delta R_\text{det}^\text{el}$ at magnetic fields below $\SI{5}{\tesla}$. This originates from the contribution of low-energy magnons, which become more dominant for increasing film thickness $t_\mathrm{m}$ \cite{Guckelhorn2022}. Additionally, we observe an oscillating behavior of $\Delta R_\text{det}^\text{el}$ at high magnetic fields in agreement with our previous experiments \cite{Guckelhorn2022}. Compared to the thin \mbox{NAOS-Fe$_2$O$_3$} film in (b), $\Delta R_\text{det}^\text{el}$ is one order of magnitude larger due to the higher density of magnonic states in thick \mbox{$\alpha$-Fe$_2$O$_3$} \cite{Guckelhorn2022}. Furthermore, $\mu_0H_\text{c}$ is shifted to a smaller magnetic field of $\SI{5.4}{\tesla}$, indicating a change in the magnetic anisotropy in thick $\alpha$-Fe$_2$O$_3$ films. 

The corresponding results of the thin and thick \mbox{AOS-Fe$_2$O$_3$} films are presented in Fig.\,\ref{fig:hanle} (d) and (e). Clearly, the maximum value of $\Delta R_\text{det}^\text{el}$ is larger than for the \mbox{NAOS-Fe$_2$O$_3$} counterparts shown in (b) and (c). Since the injector-detector distances of the respective NAOS- and \mbox{AOS-Fe$_2$O$_3$} samples are comparable, the increased $\Delta R_\text{det}^\text{el}(\mu_0H_\text{c})$ for \mbox{AOS-Fe$_2$O$_3$} cannot be explained by a decrease in $d$. Fits to Eq.\,\eqref{eq: hanle} (see Fig.\,\ref{fig:hanle} (b), (d)) reveal a larger magnon spin decay length for the device with $d=\SI{750}{\nano\meter}$ on \mbox{AOS-Fe$_2$O$_3$} than for the \mbox{NAOS-Fe$_2$O$_3$} ($d=\SI{700}{\nano\meter}$) sample. For the devices with $d=\SI{1000}{\nano\meter}$, the extracted fit parameters suggest that the increased $\Delta R_\text{det}^\text{el}(\mu_0H_\text{c})$ for \mbox{AOS-Fe$_2$O$_3$} originates from an increased factor $j_\mathrm{s0}/\chi$. As the measured spin Hall magnetoresistance (SMR) at the injector of all \mbox{$\alpha$-Fe$_2$O$_3$} samples is in the same order of magnitude (see SM), the larger $\Delta R_\text{det}^\text{el}(\mu_0H_\text{c})$ for \mbox{AOS-Fe$_2$O$_3$} cannot be attributed to an enhanced spin current transparency of the $\alpha$-Fe$_2$O$_3$/Pt interfaces, as it would also affect $j_\mathrm{s0}$. Instead, the prefactor $\chi$ is reduced, which describes a change in the magnon density of states possibly due to a change in the magnetic anisotropy of \mbox{$\alpha$-Fe$_2$O$_3$}. Thus, a larger $l_\mathrm{m}$ as well as a smaller $\chi$ can explain the larger magnon Hanle peak signal for \mbox{AOS-Fe$_2$O$_3$} films. In addition to the increase in $\Delta R_\text{det}^\text{el}(\mu_0H_\text{c})$, we observe a narrowing of the magnon Hanle peaks for the thin \mbox{AOS-Fe$_2$O$_3$} film compared to the corresponding \mbox{NAOS-Fe$_2$O$_3$} film (cf.\,Fig.\,\ref{fig:hanle} (b) and (d)). As the injector-detector distances of the respective NAOS- and \mbox{AOS-Fe$_2$O$_3$} samples are comparable with each other, the fits to Eq.\,\eqref{eq: hanle} suggest an increase in the fit parameter $c_2$ describing the magnetic field dependent contribution to the pseudofield $\omega=-c_1+c_2H$ and thus a change in the frequency of the pseudospin precession around its equilibrium pseudofield (see SM for details on the fitting procedure and extracted fit parameters)\cite{Wimmer2020}.

Moreover, we observe a decrease in the compensation field $\mu_0H_\mathrm{c}$ to $\SI{7.4}{\tesla}$ and $\SI{4.2}{\tesla}$ for devices on thin and thick \mbox{AOS-Fe$_2$O$_3$} films, respectively (see vertical dashed lines in Fig.\,\ref{fig:hanle} (d), (e)). At the compensation field $\mu_0H_\mathrm{c}$, the external magnetic field compensates the easy-plane anisotropy field in \mbox{$\alpha$-Fe$_2$O$_3$} resulting in a vanishing pseudofield \cite{Wimmer2020}
\begin{equation}
    \hbar \omega = \hbar\omega_\mathrm{an}-\mu_0m_\mathrm{net}H_\mathrm{DMI} = \hbar\tilde{\omega}_\mathrm{an}-\mu_0\tilde{m}H = 0
\end{equation}
with the easy-plane anisotropy $\hbar\omega_\mathrm{an}$ and the reduced anisotropy energy $\hbar\tilde{\omega}_\mathrm{an}$. The net magnetic moment is given by $m_\mathrm{net}$ and the DMI field by $H_\mathrm{DMI}$, which is also taken into account by the magnetic moment $\tilde{m}$. 
The shift in $\mu_0H_\mathrm{c}$ for devices on \mbox{AOS-Fe$_2$O$_3$} films is therefore associated with a change in the magnetic anisotropy in \mbox{AOS-Fe$_2$O$_3$} with respect to the \mbox{NAOS-Fe$_2$O$_3$} films (cf.\,Morin transitions in Fig.\,\ref{fig:xrd}(c) and (d)). The decrease in $\mu_0H_\mathrm{c}$ is in agreement with the increase of $T_\mathrm{M}$ as we expect a reduced strength of the easy-plane anisotropy 
for the same $T$. Overall, we are able to tune the magnetic anisotropy and the magnon Hanle effect in \mbox{$\alpha$-Fe$_2$O$_3$} by adding atomic oxygen during the deposition process.

\begin{figure}
\includegraphics{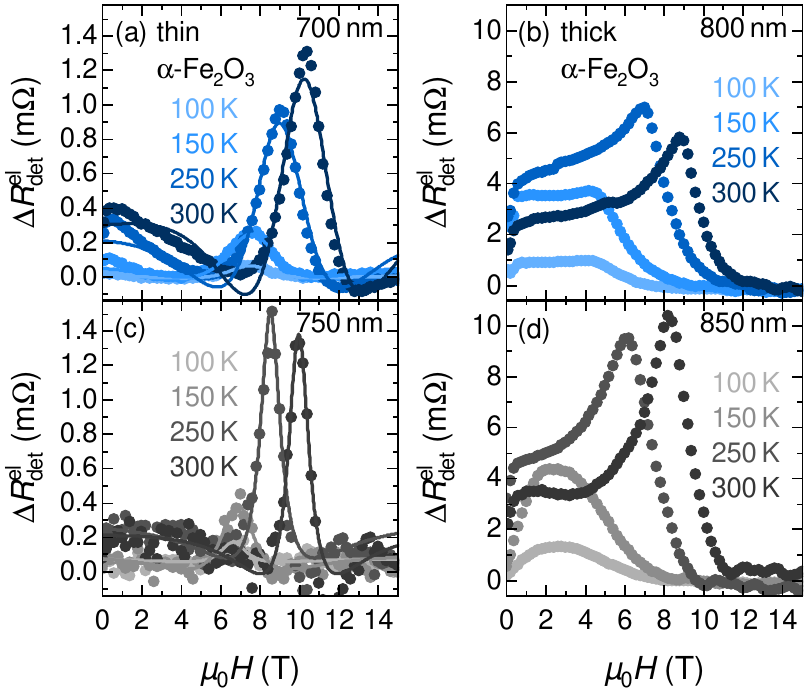}
\caption{\label{fig:hanleT} Magnetic field dependence of the electrically induced magnon spin signal amplitude $\Delta R_\text{det}^\text{el}$ at different temperatures for (a) a $d=\SI{700}{\nano\meter}$-device on a thin ($\SI{15}{\nano \meter}$) NAOS-Fe$_2$O$_3$ film, (b) a $d=\SI{800}{\nano\meter}$-device on a thick ($\SI{103}{\nano \meter}$) NAOS-Fe$_2$O$_3$ film, (c) a $d=\SI{750}{\nano\meter}$-device on a thin ($\SI{19}{\nano \meter}$) \mbox{AOS-Fe$_2$O$_3$} as well as (d) a $d=\SI{850}{\nano\meter}$-device on a thick ($\SI{89}{\nano \meter}$) AOS-Fe$_2$O$_3$ film. The solid lines in (a) and (c) are fits to Eq.\,\eqref{eq: hanle}.}
\end{figure}

To further investigate the differences in $\Delta R_\text{det}^\text{el}$ between devices on \mbox{NAOS-Fe$_2$O$_3$} and \mbox{AOS-Fe$_2$O$_3$} films, we measure $\Delta R_\text{det}^\text{el}(H)$ at different temperatures ranging from $\SI{100}{\kelvin}$ to $\SI{300}{\kelvin}$. 
The results are given in Fig.\,\ref{fig:hanleT}, where we focus on datasets for devices with fixed injector-detector distances for each \mbox{$\alpha$-Fe$_2$O$_3$} film. With increasing $T$, $\mu_0H_\text{c}$ clearly shifts to higher magnetic fields as the magnetic anisotropy in \mbox{$\alpha$-Fe$_2$O$_3$} is temperature dependent \cite{Besser1967}. Moreover, $\Delta R_\text{det}^\text{el}$ at $\mu_0H_\text{c}$ first increases with increasing $T$ due to an increased amount of thermally occupied magnon states \cite{Gonnenwein2015,Cornelissen2016}. In (b) and (c), however, the maximum of $\Delta R_\text{det}^\text{el}$ decreases again for $T>\SI{250}{\kelvin}$ as a result of an increase in magnon scattering processes \cite{Han2020}. For the thin \mbox{AOS-Fe$_2$O$_3$} film presented in Fig.\,\ref{fig:hanleT} (c), no significant changes in $\Delta R_\text{det}^\text{el}(\mu_0H)$ across the Morin transition temperature  $T_\mathrm{M} = \SI{125}{\kelvin}$ can be observed due to the low $T_\mathrm{M}$, which further decreases for $\mu_0H > 0$. In case of devices on the thick films, distinct differences in $\Delta R_\text{det}^\text{el}$ between devices on \mbox{NAOS-Fe$_2$O$_3$} (Fig.\,\ref{fig:hanleT} (b)) and on \mbox{AOS-Fe$_2$O$_3$} films (Fig.\,\ref{fig:hanleT} (d)) appear at $T \leq \SI{150}{\kelvin}$. For $\mu_0H<\SI{4}{\tesla}$, the plateau in $\Delta R_\text{det}^\text{el}$ originating from the contribution of low-energy magnons disappears in AOS-Fe$_2$O$_3$ and a peak-like behavior becomes visible instead (see Fig.\,\ref{fig:hanleT} (d)). From the linear fit to $T_\mathrm{M}(\mu_0H)$ in the inset in Fig.\,\ref{fig:xrd} (d), we can conclude that at $\SI{150}{\kelvin}$ and at $\SI{100}{\kelvin}$ the \mbox{AOS-Fe$_2$O$_3$} film is partially in the easy-axis phase for magnetic fields up to $\SI{3}{\tesla}$ and $\SI{6}{\tesla}$, respectively. The peak-like behavior of $\Delta R_\text{det}^\text{el}$ is therefore in agreement with measurements in easy-axis \mbox{$\alpha$-Fe$_2$O$_3$}, where it is attributed to a spin reorientation induced by the DMI \cite{Ross2020}.

\begin{figure}
\includegraphics{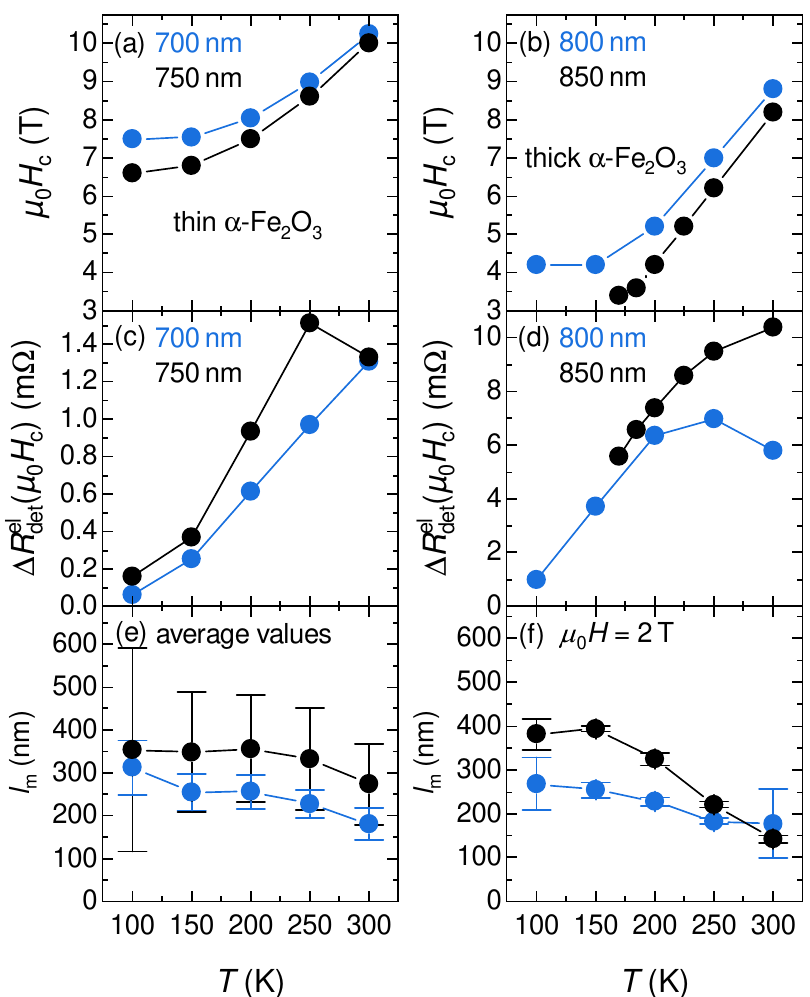}
\caption{\label{fig:parameters} (a), (b) Compensation field $\mu_0H_\mathrm{c}$, (c), (d) maximum detected spin signal amplitude $\Delta R_\text{det}^\text{el}$($\mu_0H_\mathrm{c}$) of the electrically excited magnons, and (e), (f) magnon spin decay length $l_\mathrm{m}$ as functions of temperature $T$. The left (right) panels correspond to thin (thick) \mbox{$\alpha$-Fe$_2$O$_3$} films and the blue (black) dots represent the NAOS (AOS)-Fe$_2$O$_3$ films. In (a)-(d), the injector-detector distances are given in the respective panels, whereas in (e) the average of $l_\mathrm{m}$ for different investigated injector-detector distances is given. In (f), $l_\mathrm{m}$ is taken at a magnetic field of $\SI{2}{\tesla}$ by the fitting procedure described in the SM. The solid lines in (a)-(f) are guides to the eye.}
\end{figure}

We extract the compensation field $\mu_0H_\mathrm{c}$ and the electrically induced magnon spin signal amplitude $\Delta R_\text{det}^\text{el}$ at $\mu_0H_\mathrm{c}$ from the magnon Hanle curves recorded at different temperatures $\SI{100}{\kelvin} \leq T \leq \SI{300}{\kelvin}$ and present both as functions of $T$ in Fig.\,\ref{fig:parameters} (a)-(d).  The compensation field exhibits the expected $T^2$ dependence for thin (a) as well as for thick \mbox{$\alpha$-Fe$_2$O$_3$} films (b) \cite{Besser1967}. Compared to the \mbox{NAOS-Fe$_2$O$_3$} films (blue symbols), the respective compensation fields of the \mbox{AOS-Fe$_2$O$_3$} films (black symbols) are reduced over the whole temperature range, but converge at higher temperatures. The thick \mbox{AOS-Fe$_2$O$_3$} film (black in Fig.\,\ref{fig:parameters} (b)) exhibits the Morin transition at sufficiently large temperatures, below which the magnon Hanle effect simply vanishes. Correspondingly, $\mu_0H_\mathrm{c}$ and $\Delta R_\text{det}^\text{el}(\mu_0H_\mathrm{c})$ can not be extracted for thick \mbox{AOS-Fe$_2$O$_3$} at temperatures below $\SI{170}{\kelvin}$.
At $\SI{100}{\kelvin} \leq T \leq \SI{300}{\kelvin}$, the amplitude $\Delta R_\text{det}^\text{el}(\mu_0H_\mathrm{c})$ shown in Fig.\,\ref{fig:parameters} (c) and (d) is larger for \mbox{AOS-Fe$_2$O$_3$} films than for \mbox{NAOS-Fe$_2$O$_3$} films despite the slightly larger injector-detector distances for the \mbox{AOS-Fe$_2$O$_3$} films. In addition, in (c) the maximum of $\Delta R_\text{det}^\text{el}(\mu_0H_\mathrm{c})$ of \mbox{AOS-Fe$_2$O$_3$} is shifted to lower $T$ and in (d) to higher $T$ compared to the \mbox{NAOS-Fe$_2$O$_3$} films. Thus, the influence of magnon scattering at higher $T$ strongly depends on the individual \mbox{$\alpha$-Fe$_2$O$_3$} film. Additional data for different $d$ are provided in the SM and confirm the observed $\Delta R_\text{det}^\text{el}(T)$ behavior at $\mu_0H_\mathrm{c}$.

To investigate differences between NAOS- and \mbox{AOS-Fe$_2$O$_3$} films in terms of the magnon spin decay length $l_\mathrm{m}$, we extract $l_\mathrm{m}$ from the fits to Eq.\,\eqref{eq: hanle} depicted in Fig.\,\ref{fig:hanle} (b) and (d) and in Fig.\,\ref{fig:hanleT} (a) and (c) for the thin \mbox{$\alpha$-Fe$_2$O$_3$} films. For the thick \mbox{$\alpha$-Fe$_2$O$_3$} films it is not possible to adequately fit $\Delta R_\text{det}^\text{el}(\mu_0H)$. Therefore, we determine the magnon spin decay length of thick \mbox{$\alpha$-Fe$_2$O$_3$} by a different fitting procedure as described in the SM. In Fig.\,\ref{fig:parameters} (e) and (f), $l_\mathrm{m}$ is plotted as a function of $T$ for thin and thick \mbox{$\alpha$-Fe$_2$O$_3$}, respectively. Hereby in (e), we average over all investigated two-terminal devices with $d$ ranging from $\SI{450}{\nano \meter}$ to $\SI{1300}{\nano \meter}$ depending on the respective thin film. In (f), $l_\mathrm{m}$ is determined at a magnetic field magnitude of $\SI{2}{T}$, at which $\Delta R_\text{det}^\text{el}(\mu_0H)$ exhibits a plateau above the Morin transition and a maximum below the Morin transition for thick \mbox{$\alpha$-Fe$_2$O$_3$}. For thin as well as for thick \mbox{$\alpha$-Fe$_2$O$_3$} films, $l_\mathrm{m}$ first increases with decreasing $T$ due to a reduction of magnon scattering processes. At lower $T$ down to $\SI{100}{\kelvin}$, $l_\mathrm{m}$ seems to saturate. In (e), the magnon spin decay length of \mbox{NAOS-Fe$_2$O$_3$} is slightly reduced compared to the one of \mbox{AOS-Fe$_2$O$_3$}. However, taking into account the given uncertainties, $l_\mathrm{m}$ is approximately the same for NAOS- and \mbox{AOS-Fe$_2$O$_3$} films. In (f), the magnon spin decay length of both \mbox{$\alpha$-Fe$_2$O$_3$} films is nearly the same for temperatures above $\SI{200}{\kelvin}$. For $T\leq\SI{200}{\kelvin}$, the magnon spin decay length of the \mbox{NAOS-Fe$_2$O$_3$} film is slightly smaller than for the \mbox{AOS-Fe$_2$O$_3$} film. As we expect a partial reduction of Fe$^{3+}$ to Fe$^{2+}$ ions in oxygen deficient \mbox{$\alpha$-Fe$_2$O$_3$} films, the \mbox{NAOS-Fe$_2$O$_3$} films could exhibit a higher density of Fe$^{2+}$ ions than the \mbox{AOS-Fe$_2$O$_3$} films, where the atomic-oxygen assisted deposition should enhance the oxygen content. Therefore, increased magnon scattering at Fe$^{2+}$ ions could reduce the magnon spin decay length in \mbox{NAOS-Fe$_2$O$_3$}. All in all, the magnon spin decay length of the four investigated \mbox{$\alpha$-Fe$_2$O$_3$} samples is in the same order of magnitude, but sightly larger for \mbox{AOS-Fe$_2$O$_3$}. Hence, a change in the magnetic anisotropy affecting the effective susceptibility $\chi$ of the \mbox{AOS-Fe$_2$O$_3$} films and the magnon spin decay length $l_\mathrm{m}$ in \mbox{AOS-Fe$_2$O$_3$} is possibly causing the increase in the amplitude of the magnon Hanle peak.

\section{Conclusion}
We fabricated \mbox{$\alpha$-Fe$_2$O$_3$} films of different thickness without (\mbox{NAOS-Fe$_2$O$_3$}) as well as with the addition of atomic oxygen during the deposition process (\mbox{AOS-Fe$_2$O$_3$}). 
We study the Morin transition in our samples, which is sensitive to the magnetic anisotropy in \mbox{$\alpha$-Fe$_2$O$_3$}. The \mbox{NAOS-Fe$_2$O$_3$} films exhibit no Morin transition down to $\SI{10}{\kelvin}$, while the \mbox{AOS-Fe$_2$O$_3$} films clearly show a Morin transition at around $T_\mathrm{M}=\SI{125}{\kelvin}$ for the $\SI{19}{\nano \meter}$ thin and at around $\SI{205}{\kelvin}$ for the $\SI{89}{\nano \meter}$ thick film. This proves that we are able to tune the magnetic anisotropy and thus the Morin transition in thin as well as in thick \mbox{$\alpha$-Fe$_2$O$_3$} films using an atomic oxygen source. Furthermore, the thick \mbox{AOS-Fe$_2$O$_3$} film reveals an additional Morin transition at $\SI{125}{\kelvin}$, which indicates the existence of a second \mbox{$\alpha$-Fe$_2$O$_3$} phase with a different magnetic anisotropy near the film-substrate interface. 

We conducted all-electrical magnon transport measurements above and below $T_\mathrm{M}$ and studied the magnon Hanle effect in our \mbox{$\alpha$-Fe$_2$O$_3$} films. Adding atomic oxygen during the growth process leads to two distinct changes: (i) an increase in the electrically induced magnon spin signal amplitude $\Delta R_\text{det}^\text{el}$ at the compensation field $\mu_0H_\mathrm{c}$ and (ii) a reduction of $\mu_0H_\mathrm{c}$. The increase in $\Delta R_\text{det}^\text{el}$ originates from an increase in the magnon spin decay length $l_\mathrm{m}$ or a decrease of the effective susceptibility $\chi$ describing a change in the magnon density of states. In addition, at $\mu_0H_\mathrm{c}$ the easy-plane anisotropy of \mbox{$\alpha$-Fe$_2$O$_3$} is compensated by the externally applied magnetic field. Therefore, the reduction of $\mu_0H_\mathrm{c}$ is consistent with the increase of $T_\mathrm{M}$ as for \mbox{AOS-Fe$_2$O$_3$} the strength of the easy-plane anisotropy is reduced compared to the strength of the easy-axis anisotropy.

In summary, our results show that a variation of the oxygen content during the deposition process as well as a variation of the film thickness alters the magnetic anisotropy of $\alpha$-Fe$_2$O$_3$. Our results provide a pathway towards a finite $T_\mathrm{M}$ in hematite films with thicknesses below \SI{100}{nm} and a new perspective on the role of magnetic anisotropy for the magnon Hanle effect.

\section*{Supplementary material}
Additional XRD data, details on the fabrication and measurement methods and all-electrical magnon transport data are provided in the supplementary material. This includes reciprocal space mappings, calculated lattice constants as well as explanations on the fitting procedures, extracted fit parameters and further temperature dependent data.

\begin{acknowledgments}
We gratefully acknowledge financial support from the Deutsche Forschungsgemeinschaft (DFG, German Research Foundation) under Germany’s Excellence Strategy – EXC-2111 – 390814868, and the Spanish Ministry for Science and Innovation – AEI Grant CEX2018-000805-M (through the “Maria de Maeztu” Programme for Units of Excellence in R\&D). This research is part of the Munich Quantum Valley, which is supported by the Bavarian state government with funds from the Hightech Agenda Bayern Plus.
\end{acknowledgments}

\section*{Author declarations}
\subsection*{Conflict Of Interest}
The authors have no conflicts to disclose.
\subsection*{Author Contributions}
\textbf{Monika Scheufele:} Data curation (lead); Formal analysis (equal); Writing - original draft (lead). \textbf{Janine Gückelhorn:} Data curation (equal); Formal analysis (supporting); Investigation (lead); Writing - review \& editing (supporting). \textbf{Matthias Opel:} Formal analysis (equal); Writing - review \& editing (equal). \textbf{Akashdeep Kamra:} Formal analysis (supporting); Writing - review \& editing (supporting). \textbf{Hans Huebl:} Writing - review \& editing (supporting). \textbf{Rudolf Gross:} Funding acquisition (equal), Project administration (equal); Writing - review \& editing (supporting). \textbf{Stephan Geprägs:} Conceptualization (equal); Data curation (supporting); Formal analysis (equal); Investigation (supporting); Project administration (equal); Supervision (lead); Writing - review \& editing (equal). \textbf{Matthias Althammer:} Conceptualization (equal); Formal analysis (equal); Funding acquisition (equal); Project administration (equal); Supervision (supporting); Writing - review \& editing (equal).

\section*{Data Availability}
The data that support the findings of this study are available from the corresponding author upon reasonable request.

\appendix

\bibliography{main}

\end{document}